\documentclass[a4paper,12pt]{article}
\usepackage{epsfig}
\usepackage{amssymb}
\usepackage{cite}

\newcommand{\bqll}{$\bar{B}_{q}\rightarrow l^+ l^-$~}
\newcommand{\bs}{$B_s\rightarrow \mu^+\mu^-$~}

\newcommand{\bsd}{$B_{s,d}\rightarrow \mu^+\mu^-$~}
\newcommand{\cbs}{{\cal B}(B_s\rightarrow\mu^+\mu^-)}

\newcommand{\cbd}{{\cal B}(B_d\rightarrow\mu^+\mu^-)}
\newcommand{\cbqll}{{\cal B}(B_q\rightarrow l^+ l^-)}
\newcommand{\gsim}{\lower.7ex\hbox{$\;\stackrel{\textstyle>}{\sim}\;$}}
\newcommand{\lsim}{\lower.7ex\hbox{$\;\stackrel{\textstyle<}{\sim}\;$}}

\newcommand{\hd}{{\bf h_d}}
\newcommand{\hu}{{\bf h_u}}
\newcommand{\hhu}{{\bf \hat{h}_u}}

\newcommand{\Eg}{{\bf \hat{E}_g}}
\newcommand{\Eu}{{\bf \hat{E}_u}}

\newcommand{\asos}{{\bf 1}}

\newcommand{\Vckm}{{\bf {V}}}
\newcommand{\Md}{{\bf \hat{M}_d}}
\newcommand{\Mu}{{\bf \hat{M}_u}} 

\newcommand{\Ro}{{\bf {R}}}

\newcommand{\Egb}{{\bf {E}_g}}
\newcommand{\Eub}{{\bf {E}_u}}

\newcommand{\be}{\begin{equation}}
\newcommand{\ee}{\end{equation}}
\newcommand{\br}{\begin{eqnarray}}
\newcommand{\er}{\end{eqnarray}}

\begin{document}
\tolerance=100000

\begin{flushright}
TUM-521/03\\[-1mm]
hep-ph/0309233\\[-1mm]
September 2003 
\end{flushright}

\bigskip

\begin{center}
{\Large \bf The Higgs Penguin and its Applications :\\[4mm] An overview  }\\[1.7cm] 
{{\large Athanasios Dedes\footnote{Permanent address after $1^{\rm st}$
October 2003: Institute for Particle Physics Phenomenology, University
of Durham, DH1 3LE, UK}} \\[0.5cm] {\em Physik Department, Technische
Universit\"at M\"unchen,\\ D--85748 Garching, Germany }}
\end{center}

\vspace*{0.8cm}\centerline{\bf ABSTRACT}
\vspace{0.1cm}\noindent{\small We review the effective Lagrangian 
of the Higgs penguin in the Standard Model and its minimal
supersymmetric extension (MSSM).  As a master application of the Higgs
penguin, we discuss in some detail the B-meson decays into a
lepton-antilepton pair. Furthermore, we explain how this can probe the
Higgs sector of the MSSM provided that some of these decays are
seen at Tevatron Run II and B-factories. Finally, we present a
complete list of observables where the Higgs penguin could be strongly
involved.  }

\vspace*{\fill}
\newpage

\setcounter{equation}{0}
\section{Prologue}
\label{sec:intro}

Glashow and Weinberg in their seminal paper~\cite{GW} pointed out that
Flavour Changing Neutral Currents (FCNC) are suppressed ``naturally''
if all the down-type quarks acquire their masses through their coupling
to the same Higgs boson doublet, say $H_d$, and all the up-type quarks
through their coupling to a second Higgs boson doublet, say
$H_u$. These two doublets {\it may} [Standard Model (SM) case] or {\it
may not} [Two Higgs doublet Model type-II (2HDM) case] be charge conjugates
of each other.  In general, if no symmetry  considerations are assumed
at tree level, a departure from the above rule leads to severe
enhancement of $K-\overline{K}$ and/or $B-\overline{B}$ mixing not
seen at the experimental data~\cite{PDG}. 
 In the Minimal Supersymmetric Standard
Model (MSSM) the Glashow-Weinberg scheme is naturally realized due to
the holomorphicity of the superpotential; Higgs FCNC processes
appear only at loop level  when the
holomorphicity is violated by finite radiative threshold corrections
due to the soft SUSY breaking interactions\cite{Banks}.  In
this brief review, we shall present the effective Lagrangian of the
Higgs boson FCNC's in the SM and MSSM, which we term ``Higgs
penguins'', and discuss their significance in B-, K-meson and
$\tau$-lepton physics.

\section{The Higgs penguin }

\begin{center}
\begin{figure}[h]
\centerline{\psfig{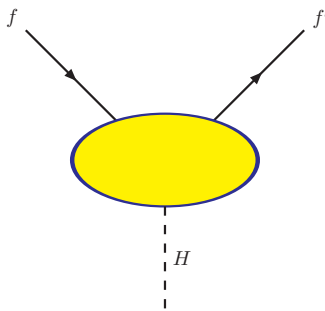}}
\caption{\it The Higgs penguin. The blob consists of loop corrections
due to Standard or Beyond the Standard Model particles.}\label{HP}
\end{figure}
\end{center}

The term ``Higgs penguin'' is used here in analogy to the 
well known Z-penguin, to denote  $H-f-f'$ loop induced 
flavour transitions, where $f,f'$ are either quarks or leptons.
For B- or K-physics experiments, the relevant Higgs penguin is the
one with down quarks $(f=d)$ in the external legs, drawn  
schematically  in Fig.{\ref{HP}}. 

\subsection{The Standard Model Higgs penguin}
\label{SM}

\begin{figure}[t]
\centerline{\psfig{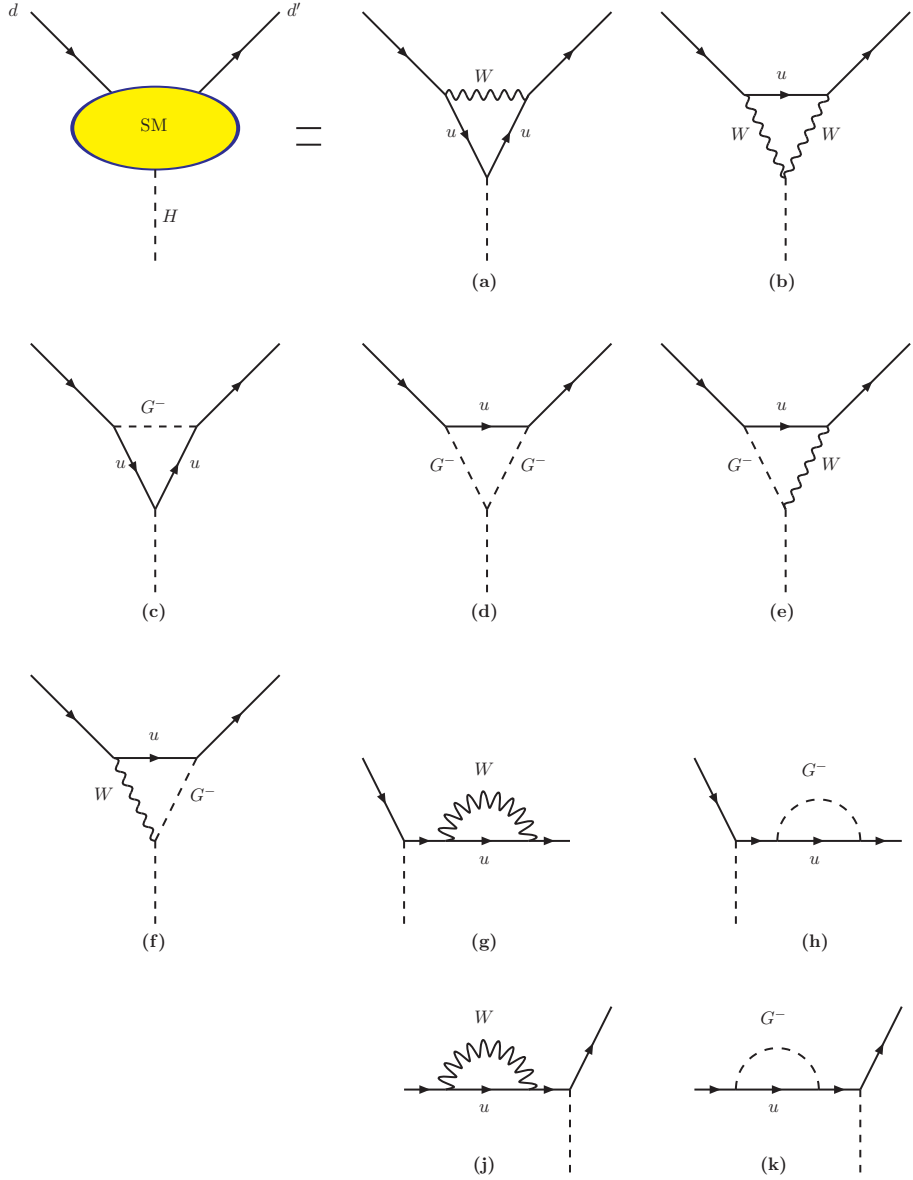}}
\caption{\it SM contributions to the  Higgs penguin. The label 
$u$ denotes all the up-type quarks $(u,c,t)$, and $G^\pm$ the charged
Goldstone modes. }\label{SMHP}
\end{figure}

The flavour changing Higgs vertex $H-d-d'$ in the SM was first calculated
in~\cite{Willey,Krawczyk}. It came rather as a surprise to
observe that the coupling of the resulting Higgs $H-d-d'$ 
penguin vertex is proportional to $m_{d(d')}/M_W$ instead of being suppressed
by $(m_{d(d')}/M_W)^3$. Denoting the up and down $3\times 3$ diagonal quark
mass matrices with $\Mu$ and $\Md$, and with $\Vckm$ the
Cabbibo-Kobayashi-Maskawa (CKM) matrix, the SM one-loop effective
Lagrangian for quark flavour changing interactions with the Higgs
boson reads
\begin{eqnarray}
  \label{SMFCNC}
{\cal L}^{\rm SM}_{H\bar{d}   d' }
\  =\ -\,  \frac{g_w}{2  M_W}\, 
H\:  \bar{d}\, \Big(\, \Md\, {{\bf g}}_{H\bar{d}  d' }^L\, P_L\ +\
{{\bf g}}_{H\bar{d}  d' }^R\, {\bf \hat{M}_{d'}} P_R\, \Big)\, d'
\,, \label{LSM}
\end{eqnarray}
where $P_{L\, (R)} = [1 -(+)\, \gamma_5 ]/2$, and
\begin{eqnarray}
  \label{SMcouplings}
{{\bf g}}_{H\bar{d} d'(d\ne d') }^L \: \!\!=\!\!
\: -\frac{3}{4}\:\frac{g_w^2}{(16\pi^2)} \:\Vckm^\dagger\: 
\frac{\Mu^2} {M_W^2}\:
\Vckm 
\;\;\;\; , \;\;\;\;  
{{\bf g}}_{H_i\bar{d} d' }^R 
\!\!&=&\!\!  \big(\,{{\bf g}}_{H_i\bar{d} d' 
  }^L\,\big)^\dagger\, .
\end{eqnarray}
The tree level flavour diagonal quark Higgs couplings are ${{\bf
g}}_{H_i\bar{d} d }^R = \big(\,{{\bf g}}_{H_i\bar{d}
d}^L\,\big)^\dagger = \asos$.  For this calculation the masses and the
momenta of the external particles have been set to zero.  This simple
result arises from the sum of 10 diagrams in $R_\xi$ gauge (6 Higgs
penguins and 4 self energy diagrams) shown in Fig.(\ref{SMHP}).
In a  general $R_\xi$ gauge, the diagrams (c),(e),(f),(g),(h),(j),(k) are
divergent.  The Glashow-Iliopoulos-Maiani (GIM) mechanism~\cite{GIM}
removes divergences from diagrams (e) and (f) aswell as from (g) and (j).
The remaining divergences in diagrams (c), (h) and (k) cancel each other
and the finite sum,  in
the case of a constant background Higgs 
field\footnote{For non zero Higgs boson mass,
the coupling ${\bf g}_{H\bar{d} d'(d\ne d') }^L$  is in general written as:
\begin{eqnarray}
{\bf g}_{H\bar{d} d'(d\ne d') }^L\: \!\!=\!\! -\frac{g_w^2}{16\pi^2} \Vckm^\dagger f(\hat{x},y) \Vckm \;,
\end{eqnarray}
with  $\hat{x}=\frac{\Mu^2} {M_W^2}$ and $y=\frac{M_H^2}{M_W^2}$, and~\cite{Krawczyk}
\begin{eqnarray}
f(\hat{x},y) = \frac{3}{4} \hat{x} + y \biggl (
-\frac{\hat{x}^3}{4 (1-\hat{x})^3} \ln \hat{x} + \frac{\hat{x}^2}{2
(1-\hat{x})^3} \ln \hat{x}  -\frac{\hat{x}^2}{8 (1-\hat{x} )^2}+\frac{3\hat{x} }{8 (1-\hat{x})^2 } \biggr ) \;.
\end{eqnarray}
Numerically $f(x\to 0 ,y)=0$, $f(4,0)=3$, $f(4,2)=3.7$, $f(4,4)=4.4$
and $f(4,100)=38.5$. Thus, for a Higgs mass around the electroweak
scale, the approximation made in deriving Eq.(\ref{SMcouplings}) is
good. However, as the Higgs mass goes over the TeV scale, the ${{\bf
g}}_{H\bar{d} d'(d\ne d') }^L$ enters into the non-perturbative
regime.  Furthermore, for applications it is the ratio
$f(\hat{x},y)/y$ which appears in the physical amplitudes. For
example, for a Higgs mass $M_H=2.6~{\rm GeV}\Rightarrow y=0.001
\Rightarrow$ $f(4,y)/y = 3000$, and $\lim_{y\to
\infty}f(4,y)/y =0.36$. }, is given in Eqs.(\ref{LSM},\ref{SMcouplings}).    
Although it seems straightforward to calculate the 10 diagrams in
Fig.(\ref{SMHP}), it took many years for gauge dependencies and
renormalization scheme ambiguities to be clarified in the
literature. The result of Refs.\cite{Willey,Krawczyk} given in
Eqs.(\ref{LSM},\ref{SMcouplings}) was finally confirmed by several
groups~\cite{Botella,Chivukula,Grinstein,Eilam,Krawczyk2,Khoze,Korner,Ferrari}
and different methods.

For $B_d$- or $B_s$-meson initial state ($d\equiv b$ and $d'\equiv d\,
~{\rm or}~\, s$), the dominant contribution in
Eq.(\ref{SMcouplings}) comes from the top quark in the loop while for
Kaon initial state ($d\equiv s$ and $d'\equiv d$) it comes
from the charm quark in the loop. Comparing the Higgs penguin
prefactor of Eq.(\ref{LSM}) and Eq.(\ref{SMcouplings}) to the
Z-boson penguin ($d\:Z_\mu\: d'$) one in Ref.~\cite{BurasLH}, 
we obtain that the former is about $m_{d}/M_W$ times
smaller than the latter, with $m_d$ being the mass of the heaviest
external quark.

\subsection{The Supersymmetric Higgs penguin}

For the construction of the  supersymmetric version of the SM Higgs
penguin two approaches have been devised: the Feynman diagrammatic
approach and the Effective Lagrangian approach.  Both have advantages
and disadvantages and it is worth describing briefly both approaches 
here. 

\subsubsection{Feynman Diagrammatic Approach}

\begin{figure}[t]
\centerline{\psfig{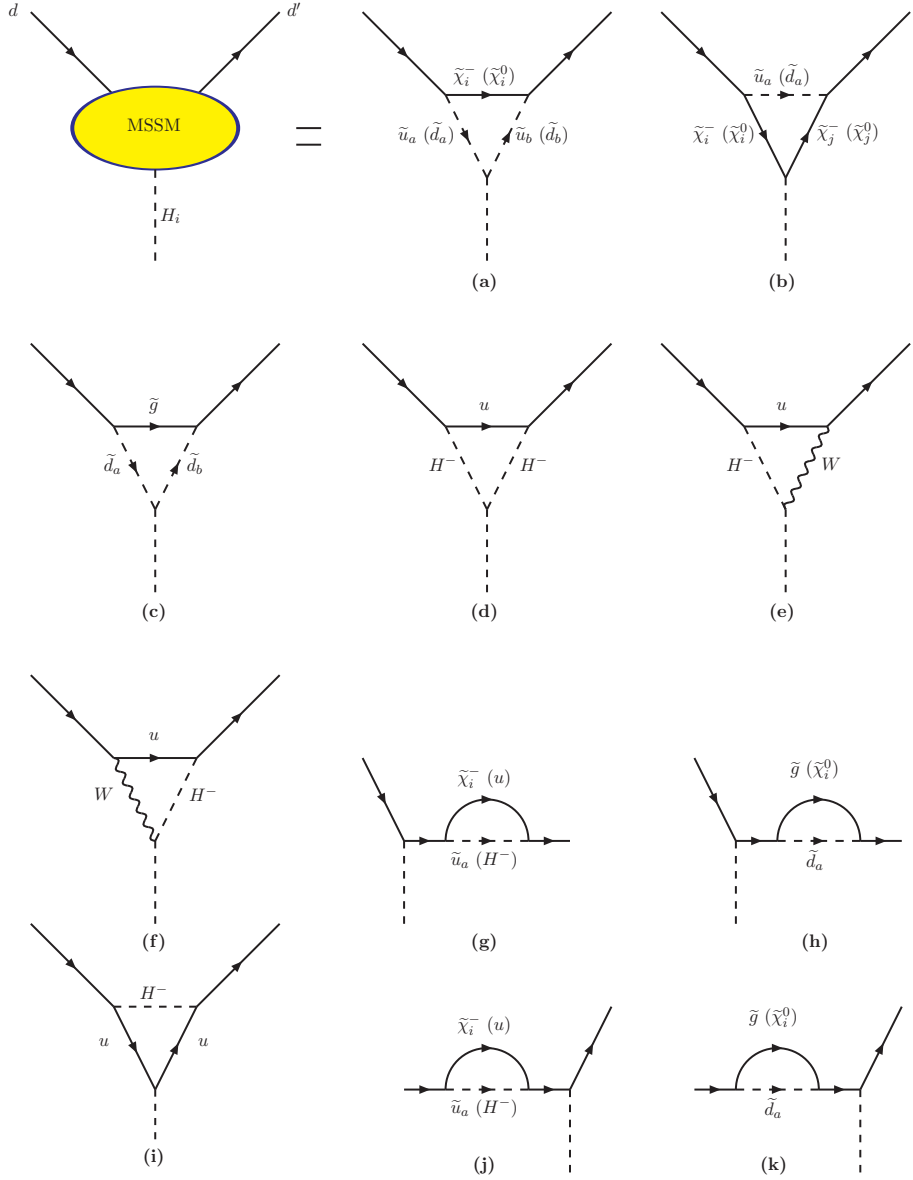}}
\caption{\it General and complete MSSM 
contributions to the Higgs penguin. The labels
$\widetilde{u},\widetilde{d}$ denote all flavours for up and down type
squarks with a=1,2 being their mass eigenstates. Charginos are labeled
with $\widetilde{\chi}_i^-, i=1,2$, neutralinos with
$\widetilde{\chi}_i^0, i=1...4$ and the gluino with
$\widetilde{g}$. $H^\pm$ is the charged Higgs boson. }\label{MSSMHP}
\end{figure}

This approach is a straightforward (but somehow tedious) calculation
of the diagrams shown in Fig.(\ref{MSSMHP}). It needs to be
supplemented with the so called $\tan\beta$-resummation procedure. The
main advantage is that within this approach effects beyond the
$SU(2)\times U(1)$ limit are automatically included. This approach has
been followed by various groups (usually through a specific  application)
\cite{Huang,Chankowski,Bobeth,Bobeth2,Dedes,Ibrahim,Curiel}. 
The 17 diagrams in Fig.(\ref{MSSMHP}) are just supersymmetrizations of
the SM diagrams in Fig.(\ref{SMHP}). To our knowledge they have not
yet been calculated in full detail~\footnote{After posting the paper,
the author was informed that in the limit of vanishing external
momenta of quarks and the Higgs particle, a complete calculation {\it
i.e.,} no expansion in $\tan\beta$,  exists in a numerical code 
in Ref.\cite{Chankowski}.}.  The latest calculation
\cite{Bobeth} is restricted only to results that have been derived in the large
$\tan\beta$ approximation with  the low $\tan\beta$ regime remaining 
unexplored although one does not expect dramatic effects for heavy (TeV)
supersymmetric particles\footnote{Non-$\tan\beta$ enhanced electroweak
corrections tend to decouple rapidly as the supersymmetric particles
approach the TeV scale.}. Some remarks on the diagrams in
Fig.(\ref{MSSMHP}) are in order here: {\it i)} the bulk of the corrections
to the Higgs penguin arise 
just from the finite part of the self energies (g,j)
with charginos and up-squarks in the
loop; they scale like $\tan^2\beta$.
They do not vanish  {\it even
(and especially) in the case of degenerate squark masses}, and the MSSM
corrections to the Higgs penguin {\it do not decouple even if the
squark masses are at multi-TeV scale}. Diagrams (d),(e),(f),(i),(g)
and (j) with the charged Higgs boson (two Higgs doublet model
contributions~
\cite{Skiba,Nierste,Bobeth} ) are also non zero but
are subdominant since they scale like $\tan\beta$. 
Ultraviolet Infinities 
of the diagrams (g,j) cancel against the vertex diagram (b).
{\it ii)} Diagrams
(a),(b),(c),(h) and (k) with neutralinos or gluinos and squarks in the
loop arise only for non degenerate squark spectrum due to
renormalization group running effects or Planck scale squark flavour
changing insertions. In this case, the gluino contributions can
compete with the chargino ones and even cancel each other (this will
be clear in a minute). The neutralino diagrams in (a,b,h,k) play
rather a subdominant role except for the case of the exact cancellation
of the gluino and chargino contributions~\cite{Bobeth2}. {\it iii)}
Within the Feynman diagrammatic approach, $\tan\beta$-resummation
effects have been added in Ref.\cite{Dedes} for all diagrams in
Fig.(\ref{MSSMHP}) apart from the gluino and neutralino ones. In
addition, SUSY CP-violating effects were taken into account in
\cite{Ibrahim}.

\subsubsection{Effective Lagrangian Approach}

\begin{figure}
\centerline{\psfig{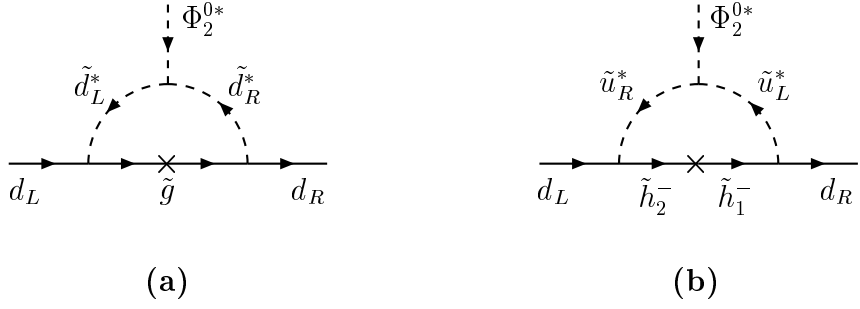}}
\caption{{\it Non-holomorphic radiative effects on the down-quark
 Yukawa couplings of the form $[\hd\: d_L \Phi_2^{0*} d_R]$ induced by
 (a) gluinos and (b) higgsinos. The Higgs field $\Phi_2^0$ is the one
 which couples to the up-type quarks in the superpotential, i.e.,
 $[\hu\: u_L \Phi_2^0 u_R]$. From Ref.\cite{DP}.}}\label{fig4}
\end{figure}

This approach allows for a simpler  derivation of the dominant
corrections to the Higgs penguin (including resummation of $\tan\beta$)
in the large $\tan\beta\gsim 40$ regime but can not easily implement
$SU(2)\times U(1)$ electroweak symmetry breaking effects 
which are relevant for small values of
$\tan\beta$.  This approach has been followed   
in Refs.~\cite{Babu,Isidori,Mizukoshi,Ambrosio,DP,Buras,Demir}.  
The general resummed Higgs penguin, which includes
CP-violating effects in the MSSM, is given by~\cite{DP}:
\begin{eqnarray}
  \label{MSSMFCNC}
{\cal L}_{H_i\bar{d}   d'}\  =\ -\,  \frac{g_w}{2  M_W}\, \sum_{i=1}^3\,
H_i\:  \bar{d}\, \Big(\, \Md\, {{\bf g}}_{H_i\bar{d}  d' }^L\, P_L\ +\
{{\bf g}}_{H_i\bar{d}  d' }^R\, {\bf \hat{M}_{d'}} P_R\, \Big)\, d' \,,
\end{eqnarray}
in the notation of~Eq.(\ref{SMFCNC}), with 
\begin{eqnarray}
  \label{MSSMcouplings}
{{\bf g}}_{H_i\bar{d} d'\:(d\ne d') }^L \!\!&=&\!\! 
\Vckm^\dagger\Ro^{-1}\Vckm\,\Biggl [ \frac{O_{1i}}{\cos\beta}\ -\
\frac{O_{2i}}{\sin\beta}
\ +\  \frac{i \: O_{3i}}{\sin \beta \: \cos \beta}\,\Biggr ]\,, \quad 
{{\bf g}}_{H_i\bar{d} d' }^R \!\!=\!\!  \big(\,{{\bf g}}_{H_i\bar{d} d'
  }^L\,\big)^\dagger\, ,
\end{eqnarray}
where $\Ro$ is the $3\times 3$ dimensional mass matrix which resums all
the $\tan\beta$ enhanced finite threshold effects. It is given by :
\begin{eqnarray}
\label{Rhat}
\Ro \ =\  \asos\: +\: \Egb \tan\beta \: +\: \Eub \tan\beta\, |\hhu|^2\:
 +\: \dots    \;,
\label{Ro}
\end{eqnarray}
where $\hhu$ is the diagonal up-Yukawa couplings, and up to small
corrections is : $\hhu = \frac{\sqrt{2}\:
\Mu}{v_2}$. Furthermore, $O_{ij}$ is a  $3\times 3$ orthogonal matrix,
which transforms the Higgs boson fields from their weak to their mass
eigenstates and  accounts for the CP-violating Higgs mixing
effects~\cite{Pilaftsis}. In addition, $\Egb$ and $\Eub$ are finite
threshold effects induced by the diagrams in Fig.(\ref{fig4}). The
ellipses in~(\ref{Ro}) denote additional (generically sub-dominant)
threshold effects like, for example, the bino-wino contribution which
corresponds to the neutralino contribution diagram  in Fig.(\ref{MSSMHP}).
 $\Egb$ and $\Eub$ are in general complex $3\times 3$ matrices but, for
 simplicity, here we will assume that they are diagonal (the reader
 may consult Ref.\cite{DP} for more details at  this point) :
\begin{eqnarray}
  \label{Eg}
\Eg &=& \asos\, 
\frac{2 \alpha_s}{3\pi}\, m_{\tilde{g}}^* \mu^*\, I(m_{\tilde{d}_L}^2,
m_{\tilde{d}_R}^2,|m_{\tilde{g}}|^2) \stackrel{M_{\rm SUSY}\gg M_W}
\longrightarrow \asos \, \frac{\alpha_s}{3\pi} \: e^{-i\:(\phi_\mu+\phi_{g})}\;, \\[2mm] \label{Eu}
\Eu &=& \asos\, \frac{1}{16\pi^2}\,\mu^* A_U^*\, I(m_{\tilde{u}_L}^2,
m_{\tilde{u}_R}^2,|\mu|^2) \stackrel{M_{\rm SUSY}\gg M_W}
\longrightarrow  \asos \,\frac{1}{32\pi^2} \: e^{-i\:(\phi_\mu+\phi_{A_U})}
\;.
\end{eqnarray}
The limit in Eqs.(\ref{Eg},\ref{Eu}) is at degenerate supersymmetric
masses, $M_{SUSY}\equiv |m_{\tilde{g}}|=|\mu|=m_{\tilde{d}_L}=
m_{\tilde{d}_R}=|A_U|=m_{\tilde{u}_L}=m_{\tilde{u}_R}$ and
$\phi_{\mu},\phi_{g},\phi_{A_U}$ are the phases of the higgsino mixing
parameter $\mu$, the gluino mass $m_{\tilde{g}}$, and the trilinear
supersymmetry breaking coupling $A_U$, respectively. The loop integral
$I(x,y,z)$ is given in~\cite{DP}. Even in this limit, $M_{\rm SUSY}\gg
M_W$ a nightmare scenario for the LHC, the Higgs penguin amplitude
remains significant and does not decouple; it depends on the various
CP-violating phases and details of the Higgs sector [mixing matrix
$(O_{ij})$] and the experimentally nearly-known quark sector.

Equations (\ref{MSSMFCNC},\ref{MSSMcouplings},\ref{Ro}) represent the Higgs
penguin amplitude in the MSSM. In order to qualitatively understand the
behaviour of the MSSM Higgs penguin, it is instructive to take the
limit of large $\tan\beta$, i.e., $1/\cos\beta \simeq \tan\beta, \;
\sin\beta \simeq 1$ and apply the limits (\ref{Eg},\ref{Eu}) in
Eqs.(\ref{MSSMFCNC},\ref{MSSMcouplings},\ref{Ro}).  Then, we obtain:
\begin{eqnarray}
  \label{AMSSMcouplings}
& &{{\bf g}}_{H_i\bar{d} d'\:(d\ne d') }^L \!\!=\!\!
-\frac{1}{4}\:\frac{g_w^2}{(16\pi^2)}\: \Vckm^\dagger \Big (\frac{\Mu^2}{M_W^2}
\Big ) \: \times \nonumber \\[2mm] & &
\frac{ \Big ( O_{1i}+i\: O_{3i} \Big ) \: 
e^{-i (\phi_\mu+\phi_{A_U})}\: \tan^2\beta}
{\Big [\asos + \frac{\alpha_s}{3\pi}\: 
e^{-i (\phi_\mu+\phi_{g})}\:\tan\beta \Big ]\:\Big [
\asos + \Big ( \frac{\alpha_s}{3\pi}\: e^{-i (\phi_\mu+\phi_{g})}\:
+ \: \frac{g_w^2}{64 \pi^2}\frac{\Mu^2}{M_W^2}\:e^{-i (\phi_\mu+\phi_{A_U})}
\Big )\:\tan\beta \Big ]}\: \Vckm \;, \nonumber \\[5mm]
& & {{\bf g}}_{H_i\bar{d} d' }^R \!\!=\!\!  \big(\,{{\bf g}}_{H_i\bar{d} d'
  }^L\,\big)^\dagger\, .
\end{eqnarray}
By looking at Eq.(\ref{AMSSMcouplings}),
we conclude the following :
\begin{itemize}
\item The MSSM Higgs penguin is enhanced by  two powers of $\tan\beta$
with respect to the SM one. Compare Eqs (\ref{AMSSMcouplings}) and
(\ref{SMcouplings}). 
\item The MSSM Higgs penguin is in general a complex number  
due to the additional supersymmetric CP-violating phases,
$\phi_{A_U}$, $\phi_{\mu}$. In the CP-invariant limit the coupling
${{\bf g}}_{H_i\bar{d} d'\:(d\ne d') }^L$ is either pure real or pure
imaginary.
\item The MSSM Higgs penguin's largest correction, Eq.(\ref{Eu}), depends
on the soft supersymmetry breaking trilinear parameter, $A_U$.  Therefore,
it is enhanced only in the case of the maximal stop-mixing case, $A_t\gsim
M_W$, which is favoured also by MSSM Higgs searches.
\item The limit $\big [\asos + \Egb \tan\beta \big ] \rightarrow 0$ in 
the master formula Eq.(\ref{MSSMcouplings}), or equivalent the approximation
$\big [\asos +
\frac{\alpha_s}{3\pi}\: e^{-i (\phi_\mu+\phi_{g})}\:
\tan\beta\big ]\rightarrow 0 $ in Eq.(\ref{AMSSMcouplings}), is attainable
and not singular.  In this limit the Higgs penguin diagram goes to a
constant value (and not to infinity) due to the GIM mechanism. This is
a result of the general resummed effective Lagrangian, equipped
in~\cite{DP}, that integrates and   improves  earlier
constructions~\cite{Babu,Isidori,Mizukoshi,Ambrosio}.
\item The MSSM Higgs penguin will dominate over all the box- or penguin-type
supersymmetric contributions 
 to the physical observables since the latter vanish due to the
super GIM mechanism in the limit of the degenerate squark
masses~\cite{Nilles,Weinbergbook}.
\item The resummation matrix  ${\bf R}$ controls the strength of
the  Higgs-mediated FCNC  effects.   For instance,  if  ${\bf   R}$ is
proportional to    unity,   then   a   kind  of   a   GIM-cancellation
mechanism      becomes   operative    and  the  Higgs-boson
contributions to all FCNC observables vanish identically.
Furthermore, not only the  top quark,  but also the other  up-type
quarks  can  give   significant  contributions to FCNC  transition
amplitudes, which are naturally included in~(\ref{MSSMFCNC}) through the
resummation matrix ${\bf R}$. All these effects are computed explicitly
within physical observables in Ref.\cite{DP}.
\item The Lagrangian in Eq.(\ref{MSSMFCNC}) captures the bulk
of the full corrections of Fig.(\ref{MSSMHP}) at large $\tan\beta
\gsim 40$ and is limited  for  supersymmetric
 breaking masses $M_{\rm SUSY}$ much heavier than the electroweak
 scale. Additional electroweak corrections (after the $SU(2)\times
 U(1)$ gauge symmetry breaking) have been calculated in~\cite{Buras}.
\end{itemize} 
We will now turn to the physical applications of the Higgs penguin.

\section{Applications}

There is a vast amount of applications one can think of. Basically,
the Higgs penguin participates in those physical observables where the
Z-penguin does so. In this section we shall discuss, in a rather
qualitative way, the effect of the Higgs penguin $[H-b-d(s)]$ on B-,
K-meson and $\tau$-lepton physics observables. The models under
consideration will be the Standard model and its minimal supersymmetric
extension, MSSM, with R-parity symmetry.

\subsection{ The Master Application : \bqll and implications }

It is rather compulsory to discuss first the $B_q$-meson ($q=d,s$)
decay to a lepton ($l=e,\mu,\tau$) pair. The decay
\bqll is mediated by the Higgs penguin, the Z-penguin and box
diagrams\cite{BurasLH}. It has been extensively discussed in the
literature, 
both in the SM~\cite{SMbmumu} and in the MSSM~\cite{MSSMbmumu} case
(for other models see Ref.\cite{Exoticbmumu}).
None of the processes \bqll have been seen so far, and the experimental
bounds on these ``very rare''  B-decays are listed in Table~1.
\begin{table}[t]\begin{center}\label{tab1}
\begin{tabular}{|c|c|c|c|}\hline
${\cal B}$(Channel) & Expt. & Bound (90\% CL) & SM prediction \\ \hline
$B_s \to e^+ e^-$ & L3~\cite{L3}
                  & $<5.4\times 10^{-5}$
                  & $(8.9 \pm 2.3 )\times 10^{-14}$ \\ \hline
$B_s \to \mu^+ \mu^-$ & CDF~\cite{CDF}
                  & $<9.5\times 10^{-7}$
                  & $(3.8\pm1.0) \times 10^{-9}$ \\ \hline
$B_s \to \tau^+ \tau^-$  & LEP~\cite{LEP}
                 & $<0.05$
                 & $(8.2\pm 2.1) \times 10^{-7}$ \\ \hline
$B_d \to e^+ e^-$ & Belle~\cite{Belle}
                  & $<1.9\times 10^{-7}$
                  & $(2.4\pm 1.4) \times 10^{-15}$\\ \hline
$B_d \to \mu^+ \mu^-$ & Belle~\cite{Belle}
                  & $<1.6\times 10^{-7}$
                  & $(1.0\pm 0.6) \times 10^{-10}$  \\ \hline
$B_d \to \tau^+ \tau^-$  & LEP~\cite{LEP}
                         & $<0.015$
                         & $(2.1 \pm 1.2)\times 10^{-8}$
                          \\
                         \hline 
\end{tabular}
\caption{{\it The experimental status and the SM 
predictions for the branching ratios ${\cal B}$(\bqll). The error in
the $B_s$ branching ratios mainly originates from the uncertainty in
$f_{B_s}=230\pm 30$ MeV~\cite{lat},
and  the  error in the $B_d$ branching ratios corresponds to
the uncertainties in $f_{B_d}=200\pm 30$ MeV~\cite{lat} and  
$|V_{td}|=0.040\pm 0.002$
  linearly added. This is the updated version of the Table 1 presented
in Ref.\cite{Dedes2}.}}
\end{center}
\end{table}
The decays \bqll are interesting probes of physics beyond the SM since
the best bound we have so far, that on $\cbs$, is three orders of
magnitude away from the SM expectation~\cite{Burasrecent}. 
In addition, the decay \bsd is
one of the experimentally favoured and  has a bright future :
Tevatron Run~II with luminosity of 2 ${\rm fb}^{-1}$ has a single
event sensitivity~\cite{Btev} of $\cbs=1.0\times 10^{-8}$
(the background from Run~I is roughly one event): 10 events at CDF 
with ${\cal L}=2~{\rm fb}^{-1}$ means a $\cbs \simeq 10^{-7}$.
A recent analysis~\cite{Kamon,Arnowitt} showed that CDF can discover 
\bs in Run IIb with an integrated luminosity of 15 ${\rm fb}^{-1}$
if $\cbs \gsim 10^{-8}$. If $\cbs$ turns out to be SM-like (see
Table~1), then only LHC will be able to measure it~\cite{Ball}.
 
Motivated by the above discussion, we are now going to see why and how
one can achieve branching ratios for the decays \bqll that could appear
soon at the Tevatron and B-factories. To this end, we shall follow the 
effective Lagrangian technique employed in Ref.~\cite{DP}.
Neglecting contributions  proportional  to the   lighter quark  masses
$m_{d,s}$, the  relevant effective Hamiltonian  for  $\Delta B=1$ FCNC
transitions, such as $b\to  q \ell^+ \ell^-$  with $q = d,s$, is given
by
\begin{eqnarray}
  \label{DB1}
H_{\rm eff}^{\Delta B=1}\ =\ -\,\frac{4\:G_F}{\sqrt{2}}\,  V_{tb}V_{tq}^*\,
\Big(\, C_S\, {\cal O}_S\ +\ C_P\, {\cal O}_P\ +\ C_{10}\, {\cal O}_{10}
\Big)\;,
\end{eqnarray}
where 
\begin{eqnarray}
{\cal O}_S &=& \frac{e^2}{16\pi^2}\, m_b\, 
(\bar{q} P_R b)\, (\bar{\ell}\ell) \;, \nonumber \\[2mm]
{\cal O}_P &=& \frac{e^2}{16\pi^2}\, m_b\, (\bar{q} P_R b)\, 
(\bar{\ell}\gamma_5 \ell) \;,\nonumber \\[2mm]
{\cal O}_{10} &=& \frac{e^2}{16\pi^2}\,  (\bar{q}\gamma^\mu P_L b)\,
 (\bar{\ell}\gamma_\mu \gamma_5 \ell) \;.
\end{eqnarray}
By making use of our master resummed Higgs penguin effective
Lagrangian Eq.(\ref{MSSMFCNC},\ref{MSSMcouplings},\ref{Ro}), the
Wilson coefficients $C_S$ and $C_P$ (in the region of large values of
$\tan\beta$) read:
\begin{eqnarray}
  \label{CSCP}
C_S &=& \frac{2 \pi m_\ell}{\alpha_{\rm em}}\, 
\frac{1}{V_{tb} V_{tq}^*}\, \sum_{i=1}^3\, \frac{g_{H_i\bar{q}b}^R\, 
g_{H_i\bar{\ell}\ell}^S}{M_{H_i}^2} \ ,\nonumber\\[3mm]
C_P &=& i\, \frac{2 \pi m_\ell}{\alpha_{\rm em}}\, \frac{1}{V_{tb} V_{tq}^*}\,
\sum_{i=1}^3\, \frac{g_{H_i\bar{q}b}^R\, 
g_{H_i\bar{\ell}\ell}^P}{M_{H_i}^2} \ ,
\end{eqnarray}
with $C_{10}=-\frac{1}{\sin^2\theta_w} (\frac{\overline{m}_t(m_t)}{167~{\rm
GeV}})^{1.55}\simeq 4.3$~\cite{SMbmumu} and $\overline{m}_t(m_t)$  the running
$\overline{MS}$ top quark mass. $M_{H_i}$ denotes the neutral Higgs boson 
masses, $M_{H_1}<M_{H_2}\lsim M_{H_3}$. 
$C_{10}$ is actually the dominant
contribution in the SM case and originates from the Z-boson penguin
and the box diagrams~\cite{BurasLH}.   
In addition, the reduced scalar and pseudoscalar Higgs couplings
to charged leptons $g_{H_i\bar{\ell}\ell}^{S,P}$ in~Eqs.(\ref{CSCP}) are
given by~\cite{DP,Apostolos2}
\begin{equation}
g_{H_i \bar{\ell} \ell}^S\ =\ \frac{O_{1i}}{\cos\beta}\ , \qquad
g_{H_i \bar{\ell} \ell}^P\ =\ -\,\tan\beta\, O_{3i} \; ,
\label{CSCPLEP}
\end{equation}
where loop vertex effects on the leptonic sector have been omitted as
being negligibly small. The reader must have already noticed that, at the
large $\tan\beta$ regime, we have $g_{H_i \bar{\ell} \ell}^S \simeq
O_{1i}\times \tan\beta$ and thus from
Eqs.(\ref{AMSSMcouplings},\ref{CSCP},\ref{CSCPLEP}) we obtain that the
scalar and pseudoscalar Wilson coefficients $C_{S,P}$ grow like
$\tan^3\beta$.
The branching ratio  for the
$\bar{B}^0_q$  meson decay  to   $\ell^+ \ell^-$  acquires the  simple
form~\cite{Bobeth,Skiba}
\begin{eqnarray}
  \label{Bll}
{\cal B}(\bar{B}^0_q \to \ell^+ \ell^-) &=& \\
&&\hspace{-2cm}
\frac{G_F^2 \alpha_{\rm em}^2}{16\pi^3}\, M_{B_q} \tau_{B_q}\, 
|V_{tb}V_{tq}^*|^2\, \sqrt{1-\frac{4 m_\ell^2}{M_{B_q}^2}}\
\Bigg[\, \Bigg(\,1-\frac{4 m_\ell^2}{M_{B_q}^2} \Bigg)\, |F^q_S|^2\ +\
|F_P^q\: +\: 2 m_\ell F_A^q|^2\, \Bigg] \;,\nonumber
\end{eqnarray}
where $\tau_{B_q}, M_{B_q}$ is the total lifetime and the mass of the
$B_q$ meson, respectively, and
\begin{eqnarray}
  \label{FSP}
F_{S,P}^q\ =\ -\,\frac{i}{2}\, M_{B_q}^2 f_{B_q}\, 
\frac{m_b}{m_b+m_q}\, C_{S,P}\;,\qquad 
F_A^q\ =\ -\,\frac{i}{2}\,f_{B_q}\, C_{10} \;.
\end{eqnarray}
$F_{S,P}$ is roughly $(m_b^2/M_W^2) \tan^3\beta$ times bigger than
$F_A$ and thus dominate for large values of $\tan\beta$. Other electroweak SUSY
box and Z-penguin contributions grow with at most two powers of
$\tan\beta$ and thus are subdominant in this region.  Numerical values
for the parameters entering Eq.(\ref{Bll},\ref{FSP}) can be found in
PDG~\cite{PDG} and in the caption of Table~(1).  The Higgs
matrix elements [$O_{ij}$] in the MSSM can be obtained from the
numerical code {\tt CPsuperH}\cite{Apostolos2}. In general,
we have $O_{11},O_{31}\ll 1$, and thus from Eqs.(\ref{CSCP}) we conclude that 
the ratio $\cbqll$ {\it depends only on the heaviest Higgs boson mass}, $M_{H_3}$.
In the simple case of
{\it equal squark masses}, and by employing 
Eqs.(\ref{AMSSMcouplings},\ref{CSCP},\ref{CSCPLEP},\ref{Bll},\ref{FSP}),
the reader can convince himself that the experimental bounds in
Table~1 are attainable in the MSSM.  In general
cases of squark non-degeneracy one should use the master formula given in
Eqs.(\ref{MSSMFCNC},\ref{MSSMcouplings}).

\subsubsection{Implications: probing the Higgs sector with \bsd
at Tevatron and B-factories}

\begin{figure}[t]
\centerline{\psfig{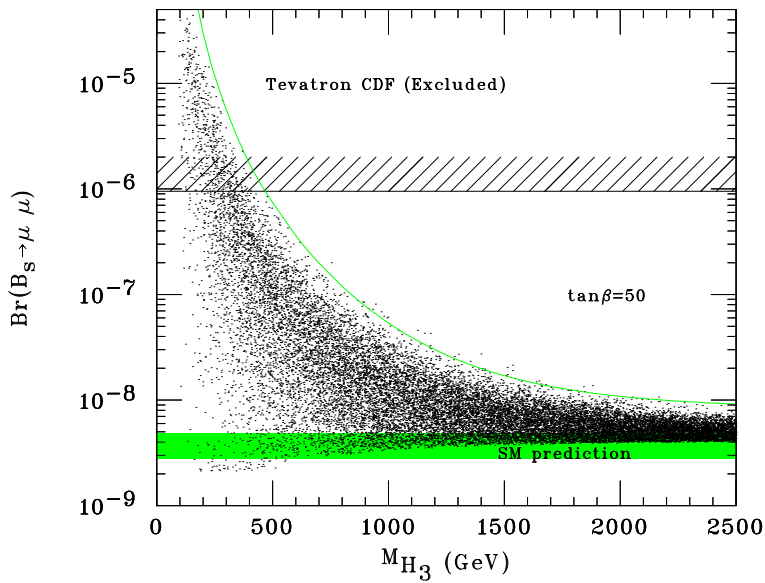}}
\caption{ {\it  The
$\cbs$ vs. the Heaviest Higgs boson mass in the MSSM for
 $\tan\beta=50$ and $m_t=175$ GeV. The shaded  area
shows the SM prediction for $\cbs$ (see Table~1). The
excluded area from Tevatron CDF~\cite{CDF} is also displayed.}}
\label{fig5}
\end{figure}
As we show, in the MSSM $\cbs$ is enhanced by six powers of
$\tan\beta$ and suppressed by four powers of the heaviest Higgs boson
mass, $M_{H_3}$. On the other hand, $\tan\beta$ has a theoretical
upper bound (around 50) coming from the perturbativity of the Yukawa
couplings up to the Grand Unification (GUT) scale. Therefore, we can
use a future possible evidence for \bs to set an upper bound on
$M_{H_3}$.  Having all the machinery at hand from the previous
section, we do this exercise here.  In order to illustrate our
argument, let us envision the following situation:
\begin{itemize}
\item Tevatron Run~II  finds \bs
\item BaBar and Belle (or a high luminosity upgraded B-factory)  
confirm the equation $\cbd \simeq  |\frac{V_{td}}{V_{ts}}|^2 \cbs $,
and no indication is found for non-minimal flavour structure
\end{itemize}
What do these {\it hypothetical} facts imply? 
 We can use a high statistics scan over the MSSM parameter space in
 the region where the soft breaking masses are below 2.5 TeV and the
 trilinear couplings below $|A|\lsim 5$ TeV.  We assume that the
 Lightest Supersymmetric particle (LSP) is stable and require it to be
 neutral. No new flavour structure other than the CKM
 is assumed. Under these assumptions, we plot in Fig.(\ref{fig5})
 $\cbs$ vs. $M_{H_3}$.  The envelope
 contour is well approximated by
\begin{eqnarray}
\cbs &=& 5\times 10^{-7}~\biggl (\frac{\tan\beta}{50}\biggr )^6 
~\biggl (\frac{550~{\rm GeV}}{M_{H_3}} \biggr )^4 
+ 8\times 10^{-9} \;. \label{bmmnum}
\end{eqnarray}
If CDF Run IIa sees 50 events for the 
\bs  [that means $\cbs \simeq 5\times 10^{-7}$]
then $M_{H_3}$ will be less than 550 GeV for {\it all} $\tan\beta$
values less than 50. The fit-equation (\ref{bmmnum}) therefore sets an
upper bound on $M_{H_3}$, a very useful result indeed.  We should note
here that various phases, or non-minimal flavour structure in the
squark sector, can alter the above result.  A more detailed analysis
is beyond the scope of this review.

\subsection{Other Applications of the Higgs Penguin}

\begin{figure}[t]
\centerline{\psfig{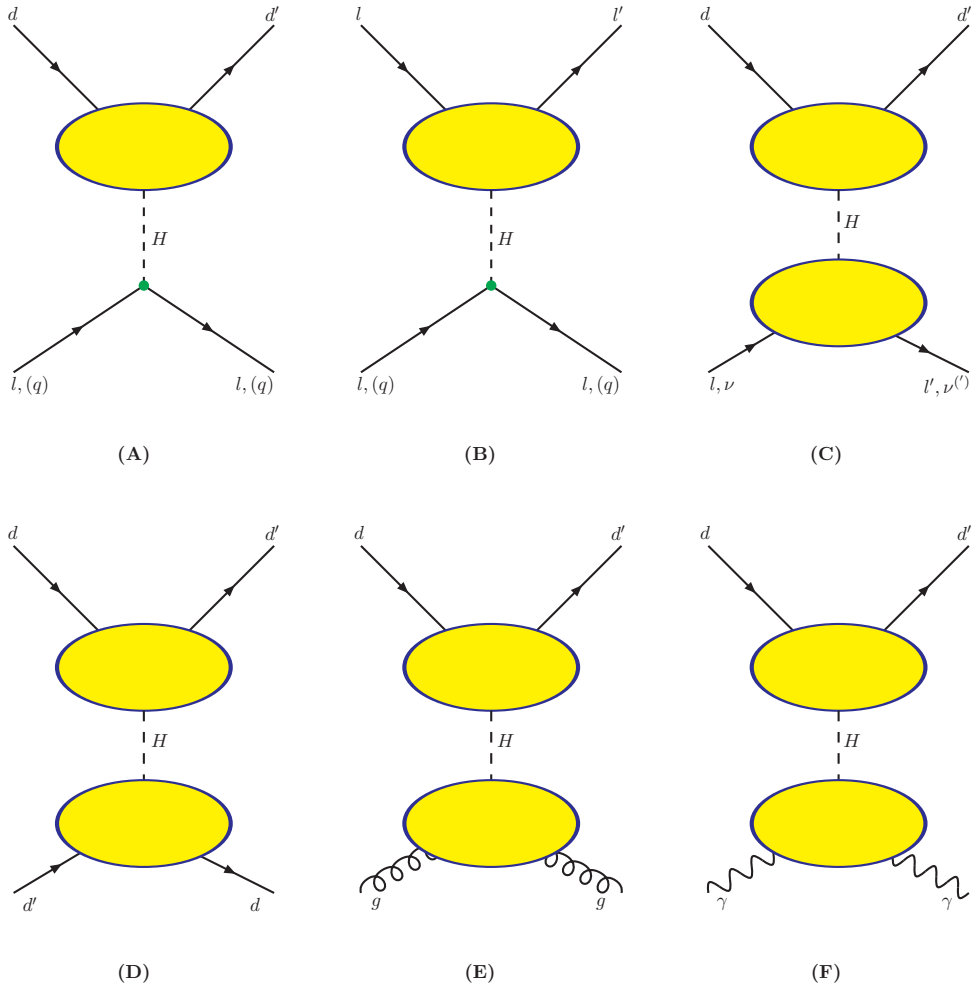}}
\caption{{\it Classification of all the applications of the Higgs penguins.
The  vertex-dot indicates that the coupling is proportional to $\tan\beta$
for the weak isospin $(-1/2)$ quarks $q=d,s,b$ and leptons $l=e,\mu,\tau$.
It is suppressed by $\cot\beta$ otherwise. The down quark Higgs penguin
(blob with $d,d'$ quarks) grows with two powers of $\tan\beta$ in
the MSSM. } }\label{appl}
\end{figure}

The Higgs penguin applications can be classified  into 
the  six categories depicted in Fig.(\ref{appl}):
\begin{itemize}
\item   
Quark single Higgs penguin decaying into a pair of leptons (l) or quarks (q)
of the same family [Fig.\ref{appl}{(A)}]. This category includes the
master application $B\to l\:l$~\cite{MSSMbmumu} that we have already
discussed in the previous section. It leads also to contributions to
the semileptonic B-decays $B\to X\:l\:l$~\cite{Bobeth,Huang2,semil}
and $K\to l\:l$. Forward-backward asymmetries in $B\to X \tau\tau$,
$B\to\pi \: l\:l$ and $K\to \pi \: l\:l$ influenced by the Higgs penguin
have been studied in~\cite{FB},~\cite{Bkll} and \cite{Kpill},
respectively, and the lepton polarization asymmetry in $B\to l\:l$ or
$K\to l\:l$ decays has been studied in~\cite{Handoko} and \cite{Kll},
respectively. The decay into a quark  pair  $d'\to d\:q\:q$ 
 leads to a number of processes  like $K\to
\pi\pi$~\cite{DP}, $B\to \pi\pi, B\to \phi K_S,B\to J/\Psi\: K, B\to
K\:\pi $ etc.  In the MSSM, the amplitude grows like $\tan^3\beta$ if
$q=d,s$ (or $l=e,\mu,\tau$),
or with $\tan\beta$ if $q=u,c$. We have seen already that the
Higgs penguin is in general complex [see for
example, Eqs.(\ref{SMFCNC},\ref{MSSMFCNC})]. As a consequence, one can
construct important CP-observables like, for example,
the leptonic CP-asymmetry~\cite{Huang3,Ibrahim,DP}.
\item  
Lepton Flavour Violating (LFV) single Higgs penguin decaying into pair
of leptons (l) or quarks (q) [Fig.\ref{appl}{(B)}].  The LFV Higgs
penguin has been calculated in the SM in Ref.\cite{Pilaftsis2}.  In
the MSSM, the method for calculating the Higgs penguin, presented in
Ref.\cite{DP} and summarized here, may be easily extended to
consistently account for charged-lepton flavour violating effective
Lagrangian. The $\tau$-decays into three leptons $\tau\to
(e,\mu)\:l\:l$~\cite{expt3mu}, or $\tau$-decays into a lepton and a
pair of quarks $\tau \to (
e,\mu) \:q\:q$,  like $\tau\to \mu\eta$,
belong to this category.  The amplitude of these decays are neutrino
mass generation mechanism dependent. For example, in the see-saw MSSM,
$\tau\to 3 l$ decays~\cite{Kolda,DER,Rossi} are enhanced by three
powers of $\tan\beta$ with the same happening in the $\tau\to l q q$
decays~\cite{Sher2}, or $\mu q\to e  q$ amplitude ($\mu-e$ conversion in the
nuclei) recently studied in~\cite{Okada}.  On the contrary, decays
with $\tau\to l u u$ grow with  only one power of $\tan\beta$.
\item
Quark-Lepton Flavour violating double Higgs penguin [Fig.\ref{appl}{(C)}].
This is nothing else than the combination of the cases (A) and (B).
The processes $B\to l\:l'$~\cite{DER}, $B\to
X \:l\:l'$ and $\tau\to (e,\mu)\:K$, etc~\cite{Sher1} belong to this category.
They start formally at 
the two-loop level and they grow with four powers of $\tan\beta$.
It is very interesting to notice that the double penguin (C) leads 
to invisible (at the detector) B-meson decays if the final particle state
is a pair of neutrinos ($\nu$) of different  flavour.
This would be  a unique signature at B-factories;
we believe that it should be further studied from the theoretical point
of view and searched for experimentally.
\item
Quark double Higgs penguin [Fig.\ref{appl}{(D)}]. Such processes lead
to contributions to $B-\bar{B}$ or $K-\bar{K}$ mixing, and in the MSSM
dominate over all other contributions for large $\tan\beta$.
Although there is no similar analysis in the SM,
following our discussion above, one expects that
the diagram (D)  is
suppressed by a factor $(m_b/M_W)^4$. In the MSSM, the dominant
contribution to the $B_{s,d}-\bar{B}_{s,d}$ mixing
is proportional to   $m_b m_{s,d}\tan^4\beta/M_{H_3}^2$~\cite{BCRS,DP,Buras}, 
and to the $K-\bar{K}$ mixing proportional to 
 $m_d m_{s,d}\tan^4\beta/M_{H_3}^2$~\cite{DP}. In addition, a complex double
penguin leads  to corrections on the CKM angle-$\beta$ originated from
soft supersymmetry breaking complex parameters~\cite{DP}.
\item
Quark-Gluon double Higgs penguin [Fig.\ref{appl}{(E)}]. This is
another class which is a combination of two penguins: the triangle
Higgs-Gluon one~(effective Lagrangian in the SM and in the MSSM can
be found in~\cite{Spira}) and the Quark Higgs penguin reviewed here.  They
contribute to decays $B\to (X)\: g\: g$ and they scale at most like
three powers of $\tan\beta$. No study of this Higgs mediated process
exists so far in the literature.
\item
Quark-Photon double Higgs penguin [Fig.\ref{appl}{(F)}]. A combination
with the photon triangle Higgs vertex - the main Higgs decay mode
search at the LHC~\cite{Spira}. It leads to processes like $B\to (X)
\gamma\gamma$ and scales at most like $\tan^3\beta$ 
when bottom quarks run in the photon triangle. No study for this amplitude
exists in the literature.
\end{itemize} 
All the above applications should  be correlated with each
other if they dominate in the corresponding processes. One example is
the correlation between $\cbs$ and $B_s-\bar{B}_s$ mass difference,
$\Delta M_s$, found in~\cite{BCRS}. This correlation is valid in a
specific scenario of squark mixing and is lost in a general
case~\cite{DP,Buras}. 
Correlations among  the full set of observables in Fig.\ref{appl}
would be  an interesting probe of the MSSM 
squark flavour structure. 
As we can see, there is still a lot of work to be done along that direction.

\section{Epilogue}

In this brief review, we discussed the effective Lagrangian of the
Higgs penguin both in the SM and in the MSSM, which are presented in
Eqs. (\ref{SMFCNC}) and (\ref{MSSMFCNC}), respectively. The SM Higgs penguin
exhibits a non-decoupling behaviour proportional to $\Vckm^\dagger\:
(\Mu^2/M_W^2)\:
\Vckm$, but it is suppressed relative to the Z-penguin by a factor of
$\sim m_d/M_W$.  The MSSM Higgs penguin exhibits also a non-decoupling 
behavior but is further enhanced by two powers of
$\tan\beta$, and dominates over the Z-penguin if $\tan\beta$ is
large. We have in addition presented the applications of the Higgs
penguin in B-meson, K-meson and $\tau$-lepton physics.  The master
application is the decay
\bqll. In the SM, the Higgs penguin contributes negligibly to $\cbqll$
if the LEP experimental bound for the Higgs boson mass is taken into
account. On the other hand in the MSSM, it is enhanced  by three orders
of magnitude mainly due to the $\tan\beta$ enhancement shown in
Eq.(\ref{AMSSMcouplings}). The B-decay mode \bsd is currently under high
search priority at Tevatron and B-factories. As we explained in section 3.1.1
one of the most interesting
aspects is the fact that a possible evidence for \bs can probe the
Higgs sector of the MSSM. Finally, we
unfolded an exhaustive list of applications in Section 3.2, some of
which are novel and not thoroughly studied even in the SM.  
We strongly believe that the Higgs penguin amplitudes deserve further 
theoretical and experimental investigation as 
their processes may soon leave their footprints 
at Hadron colliders and B-factories.

\vspace*{0.5cm}
{\bf Acknowledgements:} I would like to thank A. Buras, P.~Chankowski, F. Kr\"uger,
A. Pilaftsis, K. Tamvakis for useful discussions and comments. Many thanks to
P. Kanti for proofreading the manuscript. I would also like to
thank the CERN Theory Division for the hospitality and financial
support. I also acknowledge support by the German
Bundesministerium f\"ur Bildung und Forschung under the contract
05HT1WOA3 and the `Deutsche Forschungsgemeinschaft' DFG Project
Bu. 706/1-2.



\begin{thebibliography}{99}
\bibitem{GW}
S.~L.~Glashow and S.~Weinberg,
Phys.\ Rev.\ D {\bf 15}, 1958 (1977).


\bibitem{PDG}
K. Hagiwara et al., Phys.\ Rev.\ D{\bf 66}, 010001 (2002)


\bibitem{Banks}
T.~Banks,
Nucl.\ Phys.\ B {\bf 303} (1988) 172;
E.~Ma,
Phys.\ Rev.\ D {\bf 39} (1989) 1922;
R.~Hempfling,
Phys.\ Rev.\ D {\bf 49} (1994) 6168;
L.~J.~Hall, R.~Rattazzi and U.~Sarid,
Phys.\ Rev.\ D {\bf 50} (1994) 7048
[arXiv:hep-ph/9306309];
T.~Blazek, S.~Raby and S.~Pokorski,
Phys.\ Rev.\ D {\bf 52} (1995) 4151
[arXiv:hep-ph/9504364];
M.~Carena, M.~Olechowski, S.~Pokorski and C.~E.~Wagner,
Nucl.\ Phys.\ B {\bf 426} (1994) 269
[arXiv:hep-ph/9402253];
F.~Borzumati, G.~R.~Farrar, N.~Polonsky and S.~Thomas,
Nucl.\ Phys.\ B {\bf 555} (1999) 53
[arXiv:hep-ph/9902443].



\bibitem{Willey}
R.~S.~Willey and H.~L.~Yu,
Phys.\ Rev.\ D {\bf 26}, 3086 (1982).

\bibitem{Krawczyk}
B.~Grzadkowski and P.~Krawczyk,
Z.\ Phys.\ C {\bf 18} (1983) 43.

\bibitem{GIM}
S.~L.~Glashow, J.~Iliopoulos and L.~Maiani,
Phys.\ Rev.\ D {\bf 2} (1970) 1285.


\bibitem{Botella}
F.~J.~Botella and C.~S.~Lim,
Phys.\ Rev.\ Lett.\  {\bf 56}, 1651 (1986);
F.~J.~Botella and C.~S.~Lim,
Phys.\ Rev.\ D {\bf 34}, 301 (1986).

\bibitem{Chivukula}
R.~S.~Chivukula and A.~V.~Manohar,
Phys.\ Lett.\ B {\bf 207}, 86 (1988)
[Erratum-ibid.\ B {\bf 217}, 568 (1989)].


\bibitem{Grinstein}
B.~Grinstein, L.~J.~Hall and L.~Randall,
Phys.\ Lett.\ B {\bf 211}, 363 (1988).

\bibitem{Eilam}
G.~Eilam and A.~Soni,
Phys.\ Lett.\ B {\bf 215}, 171 (1988).



\bibitem{Krawczyk2}
The full calculation of the Higgs penguin in the SM without restricting
to zero external particle  momenta  has been presented in 
P.~Krawczyk,
Z.\ Phys.\ C {\bf 44}, 509 (1989) and in 
B.~Haeri, A.~Soni and G.~Eilam,
Phys.\ Rev.\ D {\bf 41} (1990) 875.




\bibitem{Khoze}
A.~A.~Johansen, V.~A.~Khoze and N.~G.~Uraltsev,
Sov.\ J.\ Nucl.\ Phys.\  {\bf 49}, 727 (1989)
[Yad.\ Fiz.\  {\bf 49}, 1174 (1989)].

\bibitem{Korner}
J.~G.~Korner, N.~Nasrallah and K.~Schilcher,
Phys.\ Rev.\ D {\bf 41}, 888 (1990).

\bibitem{Ferrari}
R.~Ferrari, A.~Le Yaouanc, L.~Oliver and J.~C.~Raynal,
Phys.\ Rev.\ D {\bf 52}, 3036 (1995).




\bibitem{BurasLH}
T.~Inami and C.~S.~Lim,
Prog.\ Theor.\ Phys.\  {\bf 65} (1981) 297
[Erratum-ibid.\  {\bf 65} (1981) 1772].
See also p.25-28 in A.~J.~Buras,
arXiv:hep-ph/9806471.



\bibitem{Choudhury}
S.~R.~Choudhury and N.~Gaur,
Phys.\ Lett.\ B {\bf 451} (1999) 86
[arXiv:hep-ph/9810307].

\bibitem{Huang}
C.~S.~Huang, W.~Liao, Q.~S.~Yan and S.~H.~Zhu,
Phys.\ Rev.\ D {\bf 63} (2001) 114021
[Erratum-ibid.\ D {\bf 64} (2001) 059902]
[arXiv:hep-ph/0006250].


\bibitem{Chankowski}
P.~H.~Chankowski and L.~Slawianowska,
Phys.\ Rev.\ D {\bf 63} (2001) 054012
[arXiv:hep-ph/0008046].


\bibitem{Bobeth}
C.~Bobeth, T.~Ewerth, F.~Kr\"uger and J.~Urban,
Phys.\ Rev.\ D {\bf 64} (2001) 074014
[arXiv:hep-ph/0104284].


\bibitem{Bobeth2}
C.~Bobeth, T.~Ewerth, F.~Kr\"uger and J.~Urban,
Phys.\ Rev.\ D {\bf 66} (2002) 074021
[arXiv:hep-ph/0204225].

\bibitem{Dedes}
A.~Dedes, H.~K.~Dreiner and U.~Nierste,
Phys.\ Rev.\ Lett.\  {\bf 87} (2001) 251804
[arXiv:hep-ph/0108037].



\bibitem{Ibrahim}
T.~Ibrahim and P.~Nath,
Phys.\ Rev.\ D {\bf 67} (2003) 016005
[arXiv:hep-ph/0208142].


\bibitem{Curiel}
A.~M.~Curiel, M.~J.~Herrero and D.~Temes,
Phys.\ Rev.\ D {\bf 67} (2003) 075008
[arXiv:hep-ph/0210335].


\bibitem{Skiba}
W.~Skiba and J.~Kalinowski,
Nucl.\ Phys.\ B {\bf 404} (1993) 3.

\bibitem{Nierste}
H.~E.~Logan and U.~Nierste,
Nucl.\ Phys.\ B {\bf 586}, 39 (2000)
[arXiv:hep-ph/0004139].





\bibitem{Babu}
K.~S.~Babu and C.~F.~Kolda,
Phys.\ Rev.\ Lett.\  {\bf 84} (2000) 228
[arXiv:hep-ph/9909476].


\bibitem{Isidori}
G.~Isidori and A.~Retico,
JHEP {\bf 0111} (2001) 001
[arXiv:hep-ph/0110121];
JHEP {\bf 0209} (2002) 063
[arXiv:hep-ph/0208159].

\bibitem{Mizukoshi}
J.~K.~Mizukoshi, X.~Tata and Y.~Wang,
Phys.\ Rev.\ D {\bf 66} (2002) 115003
[arXiv:hep-ph/0208078].



\bibitem{Ambrosio}
G.~D'Ambrosio, G.~F.~Giudice, G.~Isidori and A.~Strumia,
Nucl.\ Phys.\ B {\bf 645} (2002) 155
[arXiv:hep-ph/0207036].


\bibitem{DP}
A.~Dedes and A.~Pilaftsis,
Phys.\ Rev.\ D {\bf 67} (2003) 015012
[arXiv:hep-ph/0209306]. 
We follow the notation and conventions of this article.


\bibitem{Buras}
A.~J.~Buras, P.~H.~Chankowski, J.~Rosiek and L.~Slawianowska,
Nucl.\ Phys.\ B {\bf 659} (2003) 3
[arXiv:hep-ph/0210145].



\bibitem{Demir}
D.~A.~Demir,
arXiv:hep-ph/0303249.



\bibitem{Pilaftsis}
 A. Pilaftsis, Phys.\  Rev.\ {\bf D58} (1998) 096010;
  Phys.\ Lett.\ {\bf B435} (1998) 88;
  A.~Pilaftsis and C.E.M.  Wagner, Nucl.\ Phys.\ {\bf B553}
  (1999) 3; D.A.  Demir, Phys.\ Rev.\ {\bf D60} (1999) 055006; S.Y.
  Choi, M.  Drees and J.S.  Lee, Phys.\ Lett.\ {\bf B481} (2000) 57;
  G.L. Kane and L.-T. Wang, Phys.\ Lett.\ {\bf B488} (2000) 383; T.
  Ibrahim and P.~Nath, Phys.\ Rev.\ {\bf D63} (2001) 035009;


\bibitem{Nilles}  See,  for  example, J.~R.~Ellis and D.~V.~Nanopoulos,
Phys.\ Lett.\ B {\bf 110}, 44 (1982);
J.F.~Donoghue,  H.P.~Nilles  and
D.~Wyler, Phys.\ Lett.\ {\bf  B128} (1983) 55; J.M.~Gerard, W.~Grimus,
A.~Raychaudhuri and G.~Zoupanos, Phys.\ Lett.\ {\bf B140} (1984) 349.

\bibitem{Weinbergbook}
See p.201-204 in    S.~Weinberg,
{\it ``The Quantum Theory Of Fields.  Vol. 3: Supersymmetry,''}, 
Cambridge,UK:  Univ. Pr. (2000) 419 p.


\bibitem{SMbmumu}
The leading order calculation of the Z-penguins and the box contributions
to the decay \bqll  in the SM was carried out by T.~Inami and C.~S.~Lim,
in Ref.\cite{BurasLH}. NLO QCD corrections
were considered in 
G.~Buchalla and A.~J.~Buras,
Nucl.\ Phys.\ B {\bf 400} (1993) 225
and later in M.~Misiak and J.~Urban,
Phys.\ Lett.\ B {\bf 451} (1999) 161
[arXiv:hep-ph/9901278].
The Higgs penguin contribution to \bsd was very popular in the eighties
because there was almost no bound on the Higgs mass. Thus the Higgs searches
were actually carried out by B-, or K-meson decays. Relevant
references are~\cite{Willey,Krawczyk,Chivukula,Grinstein,Eilam}
and J.~F.~Gunion, H.~E.~Haber, G.~L.~Kane and S.~Dawson,
{\it ``The Higgs Hunter's Guide''}, SCIPP-89/13.






\bibitem{MSSMbmumu}
There is a vast number of articles calculating the Higgs penguin
contributions to the $\cbqll$ in the 
2HDM~\cite{Skiba,Nierste,Huang,Bobeth,Huang2} and 
general MSSM~\cite{Choudhury,Babu,Huang,Chankowski,Bobeth,Bobeth2,
Isidori,Ambrosio,DP,Buras,Huang2}. mSUGRA and mGMSB/mAMSB 
results for the $\cbqll$ appeared 
in~\cite{Dedes,Arnowitt,Mizukoshi,Dedes2,Ibrahim,Baek} and  
 in~\cite{Baek}, respectively.
SO(10) GUT predictions for the  $\cbs$  have been discussed 
in~\cite{Maxim,Dedes,Raby,Blazek}.
Gluino loop NLO  corrections to \bsd have been calculated
in~\cite{QCDMSSM}.


\bibitem{Dedes2}
A.~Dedes, H.~K.~Dreiner, U.~Nierste and P.~Richardson,
arXiv:hep-ph/0207026.


\bibitem{Huang2}
C.~S.~Huang and X.~H.~Wu,
Nucl.\ Phys.\ B {\bf 657}, 304 (2003)
[arXiv:hep-ph/0212220].


\bibitem{Burasrecent}
One can further reduce the uncertainties for the SM expectation in the
$\cbd$ by using the experimental value of $\Delta M_{d}$. With the
same method one can reduce uncertainties in the SM prediction of
$\cbs$ by using $\Delta M_{s}$ when of course this is experimentally
known. This method has been proposed in A.~J.~Buras, Phys.\ Lett.\ B {\bf
566} (2003) 115 [arXiv:hep-ph/0303060].




\bibitem{Arnowitt}
R.~Arnowitt, B.~Dutta, T.~Kamon and M.~Tanaka,
Phys.\ Lett.\ B {\bf 538} (2002) 121
[arXiv:hep-ph/0203069].

\bibitem{Baek}
S.~Baek, P.~Ko and W.~Y.~Song,
Phys.\ Rev.\ Lett.\  {\bf 89} (2002) 271801
[arXiv:hep-ph/0205259];
JHEP {\bf 0303} (2003) 054
[arXiv:hep-ph/0208112].


\bibitem{Maxim}
C.~Hamzaoui and M.~Pospelov,
Phys.\ Rev.\ D {\bf 60} (1999) 036003 \,
[arXiv:hep-ph/9901363].

\bibitem{Raby}
R.~Dermisek, S.~Raby, L.~Roszkowski and R.~Ruiz De Austri,
JHEP {\bf 0304} (2003) 037
[arXiv:hep-ph/0304101].



\bibitem{Blazek}
T.~Blazek, S.~F.~King and J.~K.~Parry,
arXiv:hep-ph/0308068.


\bibitem{QCDMSSM}
C.~Bobeth, A.~J.~Buras, F.~Kr\"uger and J.~Urban,
Nucl.\ Phys.\ B {\bf 630}, 87 (2002)
[arXiv:hep-ph/0112305].


\bibitem{Exoticbmumu}
If the R-parity symmetry is replaced by a Baryon parity symmetry 
in the MSSM, 
then \bqll can appear at tree level. 
See for instance, 
J.~H.~Jang, J.~K.~Kim and J.~S.~Lee,
Phys.\ Rev.\ D {\bf 55} (1997) 7296
[arXiv:hep-ph/9701283];
A.~G.~Akeroyd and S.~Recksiegel,
arXiv:hep-ph/0209252;
D.~Guetta, J.~M.~Mira and E.~Nardi,
Phys.\ Rev.\ D {\bf 59} (1999) 034019
[arXiv:hep-ph/9806359].
For the  calculation of the  \bqll in extended technicolor models, see:
L.~Randall and R.~Sundrum,
Phys.\ Lett.\ B {\bf 312} (1993) 148
[arXiv:hep-ph/9305289];
Z.~h.~Xiong and J.~M.~Yang,
Phys.\ Lett.\ B {\bf 546}, 221 (2002)
[arXiv:hep-ph/0208147].
For the  calculation of the   
\bqll in models with extra dimensions, see:
A.~J.~Buras, M.~Spranger and A.~Weiler,
Nucl.\ Phys.\ B {\bf 660} (2003) 225
[arXiv:hep-ph/0212143];
P.~Dey and G.~Bhattacharyya,
arXiv:hep-ph/0309110.








\bibitem{L3}
W.~Adam {\it et al.}  [DELPHI Collaboration],
Z.\ Phys.\ C {\bf 72} (1996) 207.

\bibitem{CDF}
F.~Azfar,
arXiv:hep-ex/0309005, {\it presented at  ``XXIII Physics in Collision'',
Zeuthen, Germany, 26-28 June 2003.}
This is a preliminary new  Run~II bound  for the channel \bs. The 95\%
CL bound reads  $\cbs < 1.2\times 10^{-6}$. It is by a  factor of two 
improved bound relative to the ``old'' CDF Run~I bound reported in
F.~Abe {\it et al.}  [CDF Collaboration],
Phys.\ Rev.\ D {\bf 57} (1998) 3811.
The improved CDF bound makes the mode \bs the most preferred one
in probing the MSSM parameter space  among all \bqll.

\bibitem{LEP}
This is a byproduct bound from LEP searches on the mode $B^-\to \tau^- \nu$
and was pointed out in 
Y.~Grossman, Z.~Ligeti and E.~Nardi,
Phys.\ Rev.\ D {\bf 55} (1997) 2768
[arXiv:hep-ph/9607473].
There is no experimental analysis for   the channel ${\cal B}(B_s\to
\tau^+\tau^-)$.  We would like to encourage our experimental colleagues
to search for this, motivated by the fact that in the supersymmetric
extensions of the Standard model this mode is enhanced by three orders
of magnitude relative to its SM prediction in Table~1.







\bibitem{Belle}
M.-C. Chang, {\it  et al},  [Belle Collaboration], 
KEK Preprint 2003-44, Belle Preprint 2003-12, arXiv:hep-ex/0309069.
For BaBar results on rare leptonic B decays, see
T.~B.~Moore,
SLAC-PUB-9705,
{\it presented at 31st International Conference on High Energy Physics (ICHEP 2002), Amsterdam, The Netherlands, 24-31 Jul 2002};
V.~Halyo,
arXiv:hep-ex/0207010.




\bibitem{lat}
The lattice values for $f_{B_q}$ are taken from 
C.~W.~Bernard,
Nucl.\ Phys.\ Proc.\ Suppl.\  {\bf 94} (2001) 159
[arXiv:hep-lat/0011064].
They are consistent with those reported recently by 
D.~Becirevic, at {\it  2nd Workshop On The CKM Unitarity Triangle 
   5-9 Apr 2003}. 


\bibitem{Btev}
K.~Anikeev {\it et al.},
arXiv:hep-ph/0201071.


\bibitem{Kamon}
T.~Kamon  [CDF Collaboration],
arXiv:hep-ex/0301019.


\bibitem{Ball}
P.~Ball {\it et al.},
arXiv:hep-ph/0003238.


\bibitem{Apostolos2}
J.~S.~Lee, A.~Pilaftsis, M.~Carena, S.~Y.~Choi, M.~Drees, J.~Ellis and
C.~E.~Wagner, arXiv:hep-ph/0307377.



\bibitem{semil}
C.~S.~Huang and Q.~S.~Yan,
Phys.\ Lett.\ B {\bf 442} (1998) 209
[arXiv:hep-ph/9803366];
C.~S.~Huang, W.~Liao and Q.~S.~Yan,
Phys.\ Rev.\ D {\bf 59} (1999) 011701
[arXiv:hep-ph/9803460];
Y.~Wang and D.~Atwood,
arXiv:hep-ph/0304248.
P.~H.~Chankowski and L.~Slawianowska,
arXiv:hep-ph/0308032.
For a recent  review on $B\to X_s l^+ l^-$ see, 
T.~Hurth,
arXiv:hep-ph/0212304.

\bibitem{FB}
A.~S.~Cornell and N.~Gaur,
arXiv:hep-ph/0308132.


\bibitem{Bkll}
S.~R.~Choudhury, A.~S.~Cornell, N.~Gaur and G.~C.~Joshi,
arXiv:hep-ph/0307276.


\bibitem{Kpill}
C.~H.~Chen, C.~Q.~Geng and I.~L.~Ho,
Phys.\ Rev.\ D {\bf 67} (2003) 074029
[arXiv:hep-ph/0302207].


\bibitem{Handoko}
L.~T.~Handoko, C.~S.~Kim and T.~Yoshikawa,
Phys.\ Rev.\ D {\bf 65} (2002) 077506
[arXiv:hep-ph/0112149].


\bibitem{Kll}
P.~Herczeg,
Phys.\ Rev.\ D {\bf 27}, 1512 (1983).
S.~R.~Choudhury, N.~Gaur and A.~Gupta,
Phys.\ Lett.\ B {\bf 482} (2000) 383
[arXiv:hep-ph/9909258].



\bibitem{Huang3}
  C.S.~Huang and W.~Liao, Phys.\ Lett.\ {\bf B525} (2002) 107; Phys.\
  Lett.\ {\bf B538} (2002) 301; P.H.~Chankowski and L.~Slawianowska,
  Acta Phys.\ Polon.\ {\bf B32} (2001) 1895.
Y.~B.~Dai, C.~S.~Huang, J.~T.~Li and W.~J.~Li,
Phys.\ Rev.\ D {\bf 67}, 096007 (2003)
[arXiv:hep-ph/0301082].


\bibitem{Pilaftsis2}
A.~Pilaftsis,
Phys.\ Lett.\ B {\bf 285} (1992) 68.


\bibitem{expt3mu}
For a recent  experimental status of $\tau$ and $B$ LFV decays, see
Y.~Yusa, H.~Hayashii, T.~Nagamine and A.~Yamaguchi  [BELLE Collaboration],
eConf {\bf C0209101} (2002) TU13
[arXiv:hep-ex/0211017].




\bibitem{Kolda}
K.~S.~Babu and C.~Kolda,
Phys.\ Rev.\ Lett.\  {\bf 89} (2002) 241802
[arXiv:hep-ph/0206310].



\bibitem{DER}
A.~Dedes, J.~R.~Ellis and M.~Raidal,
Phys.\ Lett.\ B {\bf 549} (2002) 159
[arXiv:hep-ph/0209207].


\bibitem{Rossi}
A.~Brignole and A.~Rossi,
Phys.\ Lett.\ B {\bf 566} (2003) 217
[arXiv:hep-ph/0304081].




\bibitem{Sher2}
M.~Sher,
Phys.\ Rev.\ D {\bf 66} (2002) 057301
[arXiv:hep-ph/0207136].



\bibitem{Okada}
R.~Kitano, M.~Koike, S.~Komine and Y.~Okada,
arXiv:hep-ph/0308021.




\bibitem{Sher1}
D.~Black, T.~Han, H.~J.~He and M.~Sher,
Phys.\ Rev.\ D {\bf 66} (2002) 053002
[arXiv:hep-ph/0206056].


\bibitem{BCRS}
A.~J.~Buras, P.~H.~Chankowski, J.~Rosiek and L.~Slawianowska,
Phys.\ Lett.\ B {\bf 546} (2002) 96
[arXiv:hep-ph/0207241].



\bibitem{Spira}
M.~Spira,
Fortsch.\ Phys.\  {\bf 46} (1998) 203
[arXiv:hep-ph/9705337].


\end{thebibliography}
\end{document}